\documentclass[11pt,a4paper]{article}
\pdfoutput=1
\usepackage{jheppub}

\usepackage{color}
\usepackage{amsmath}
\usepackage{pifont}

\usepackage[utf8]{inputenc}

\usepackage{verbatim}   
\usepackage{subfigure}

\usepackage{amsfonts}
\usepackage{amssymb}
\usepackage{mathrsfs}
\usepackage{graphicx}
\usepackage{multirow}
 \usepackage{slashed}

\newcommand{\MeV}{{\, {\rm MeV}}}
\newcommand{\GeV}{{\, {\rm GeV}}}
\newcommand{\TeV}{{\, {\rm TeV}}}

\def\beq{\begin{equation}}
\def\eeq{\end{equation}}
\def\bea{\begin{eqnarray}}
\def\eea{\end{eqnarray}}
\def\bitem{\begin{itemize}}
\def\eitem{\end{itemize}}
\newcommand{\bec}{\begin{center}}
\newcommand{\eec}{\end{center}}
\newcommand{\ba}{\begin{array}}
\newcommand{\ea}{\end{array}}

\def\OO{\mathcal{O}}

\begin{document}

\title{Non-perturbative reheating and Nnaturalness} 
\author[a]{Edward Hardy}
\emailAdd{ehardy@liverpool.ac.uk}
\affiliation[a]{Department of Mathematical Sciences, University of Liverpool, Liverpool L69 7ZL, United Kingdom}

\abstract{
We study models in which reheating happens only through non-perturbative processes. The energy transferred can be exponentially suppressed unless the inflaton is coupled to a particle with a parametrically small mass. Additionally, in some models a light scalar with a negative mass squared parameter leads to much more efficient reheating than one with a positive mass squared of the same magnitude. If a theory contains many sectors similar to the Standard Model coupled to the inflaton via their Higgses, such dynamics can realise the Nnaturalness solution to the hierarchy problem. 
A sector containing a light Higgs with a non-zero vacuum expectation value is dominantly reheated and there is little energy transferred to the other sectors, consistent with cosmological constraints.
 The inflaton must decouple from other particles and have a flat potential at large field values, in which case the visible sector UV cutoff can be raised to $10~\TeV$ in a simple model.
}

\maketitle


\section{Introduction} \label{sec:1}

Conventional approaches to the Electroweak (EW) hierarchy problem such as softly broken supersymmetry are under strong pressure from collider constraints. The simplest implementations now require at least moderate fine tuning, and even with more involved model building obtaining a fully natural theory is hard (e.g. \cite{Bellazzini:2014yua,Buckley:2016tbs}). Another possibility is that a small EW scale is selected anthropically \cite{Agrawal:1997gf,ArkaniHamed:2004fb}. While plausible given the cosmological constant problem and its potential landscape solution \cite{Weinberg:1987dv} this too has challenges. First, it is unclear how essential a low EW scale is for life if the Standard Model (SM) Lagrangian parameters are allowed to vary. Even if the light fermion masses are fixed by requiring nuclei and atoms, the distribution of Yukawa couplings must still be assumed  \cite{Hall:2007ja,Donoghue:2009me,Donoghue:2016tjk}. Second, a small EW scale could have been realised through technicolour \cite{Susskind:1978ms}, but instead the landscape must prefer a fine tuned Higgs. Unlike low scale supersymmetry, which might be disfavoured due to its connection to moduli stabilisation in string theory \cite{Denef:2004cf}, there is no theoretical reason to think that technicolour models would be rare. Although our present understanding of the string landscape is limited, extra gauge groups seem to be common  \cite{Gmeiner:2005vz,Grassi:2014zxa,Taylor:2015xtz} and it is known from QCD that these can break EW symmetry.\footnote{If there is a mini-split supersymmetry spectrum \cite{Arvanitaki:2012ps} with only a moderate tuning then this issue is less worrying than if the Higgs mass is tuned all the way from the Planck Scale.}

As a result it is interesting to explore other potential solutions to the hierarchy problem, even if the models involved require unusual or theoretically questionable features. An interesting possibility is that a low EW scale might be the result of dynamics during cosmological history. An example of this is provided by relaxion models \cite{Graham:2015cka}. In these there are many vacua with varying EW scales, and the evolution of a light scalar (the relaxion) during inflation selects a vacuum with a small negative Higgs mass squared parameter. 
Viable models usually need an extremely large number of e-folds of inflation and a relaxion potential that is periodic on a much smaller scale than its overall field range (although the former might be avoided in variations, for example \cite{Hardy:2015laa,Hook:2016mqo,Patil:2015oxa}).

An alternative proposal, Nnaturalness \cite{Arkani-Hamed:2016rle}, is that there may be a large number $N$ of sectors, all resembling the SM but with different Higgs mass squared parameters. Among these there will typically be sectors that happen to have a Higgs with a mass far below the UV cutoff, but which are otherwise undistinguished. Assuming an approximately flat distribution on the Higgs mass squared parameter, the lightest Higgs is expected to have mass $\sim \Lambda/\sqrt{N}$,  where $\Lambda$ is the UV cutoff of the theory. Then, if there are dynamics such that the sector containing the lightest Higgs with a non-zero EW vacuum expectation value (VEV) is dominantly reheated after inflation, the observed smallness of the EW scale is explained. In the original model, this was accomplished by introducing a light but technically natural reheaton field, of mass $\lesssim \Lambda/\sqrt{N}$. With appropriate model building, reheating is suppressed by $1/m_h^4$ and $1/m_h^2$ for sectors with a physical Higgs mass $m_h$ and zero and non-zero Higgs VEVs respectively \cite{Arkani-Hamed:2016rle}.

In this paper we study theories in which reheating only happens through non-perturbative processes. This is the case if the inflaton's potential is such that it does not oscillate around a minimum, but instead runs away towards infinity after inflation \cite{Felder:1998vq,Felder:1999pv}.\footnote{Unlike \cite{Kofman:2004yc}, in the models we study backreaction on the inflaton's motion from the produced states, which could lead to it being trapped, is negligible.}
 The efficiency of non-perturbative reheating can be exponentially suppressed when the evolution of the system is close to adiabatic \cite{Kofman:1994rk,Kofman:1997yn}. In particular, if the inflaton is coupled to a particle with mass $m$ changing at rate $\dot{m}$ due to the motion of the inflaton, the system is adiabatic when $\dot{m}/m^2 \lesssim 1$. Consequently, such models can realise Nnaturalness: if the $N$ sectors are coupled to the inflaton via their Higgses, very little energy is transferred to those without a light Higgs. It is technically natural for $\dot{m}$ to be small due to a small coupling constant, and the dimensionful parameters in the inflaton potential can all be of order the cutoff.\footnote{The possibility that non-oscillating models could lead to sectors having different temperatures after reheating was discussed in \cite{Hardy:2017wkr}.} Further, depending on a choice of sign in the inflaton couplings, sectors with a negative Higgs mass squared parameter can be reheated much more efficiently than those for which this is positive.

For simplicity we consider a toy version of the SM without gauge bosons or fermions, neglecting the possibility of non-perturbative energy transfer to these. As we will discuss, it is likely that such processes will only be significant for reheating to sectors with small non-zero Higgs VEVs. Also the inflaton potential and couplings must be unusual, and we do not construct a complete model with UV dynamics leading to couplings of the required form. Despite these limitations, the exponential dependence of the energy transfer on a sector's Higgs mass is potentially phenomenologically interesting. For example the effective number of extra relativistic degrees of freedom due to the additional sectors is easily compatible with current constraints without tuning, although it may remain within the reach of future observations. In the models we consider the visible sector cutoff cannot be raised above $\sim 10~\TeV$ and reheating above the scale of BBN is also challenging, although it might be possible to relax these in more complete theories. A summary of our results may be found in Figure~\ref{fig:2}, which shows the energy transferred to a sector as a function of the mass of its Higgs for an inflaton with couplings given by Eq.~\eqref{eq:infcou}.

In Section~\ref{sec:2} we study non-perturbative particle production for arbitrary inflaton couplings, and in Section~\ref{sec:3} show how this can lead to models of Nnaturalness. Subsequently we consider the phenomenological constraints in Section~\ref{sec:4}, and discuss our results  in Section~\ref{sec:5}.

\section{Non-perturbative reheating} \label{sec:2}

If the inflaton is relatively strongly coupled to other particles, non-perturbative energy transfer can play an important role in reheating  \cite{Traschen:1990sw,Kofman:1994rk,Shtanov:1994ce,Kofman:1997yn}. Usually this happens if the inflaton interacts through relevant or marginal operators with coupling constants not too much smaller than $\OO\left(1\right)$. In conventional models in which inflation ends with the inflaton $\phi$ oscillating around a minimum of its potential, the non-perturbative dynamics takes the form of bursts of explosive particle production as the inflaton moves through a region in field space where total mass of its decay product is small. At this point there is a sharp change in the decay product's occupation number which becomes momentarily undefined as the system evolves non-adiabatically \cite{Kofman:1997yn}.  
The rate of energy transfer is Bose enhanced since the inflaton oscillates through the critical part of field space many times, leading to an exponentially growing rate of particle production.\footnote{The dynamics described are known as broad resonance. In a non-expanding universe significant energy transfer is possible even if the adiabatically condition is not violated, called the narrow resonance regime. In this case particle production happens steadily over the whole of the inflaton oscillation period, with the suppressed rate compensated by the  exponentially growing Bose enhancement factor. However in an expanding universe the produced particles are redshifted out of the resonance band, and typically the energy transferred through these dynamics is not significant at early times (although they may play a role in the complex out of equilibrium system left over after a period of broad resonance finishes).} A significant fraction of the total inflaton energy can be transferred through this process, known as preheating, at early times long before perturbative inflaton decays are relevant. 

However preheating terminates with an order 1 fraction of the system's energy still remaining in the inflaton sector, both in the form of a not fully depleted zero mode and a population of higher momentum inflaton states \cite{Kofman:1997yn}. This happens because the produced particles induce an effective mass for the inflaton (and potentially themselves) moving the system out of the resonance regime, and also scatter off the inflaton zero mode creating finite momentum inflaton quanta. Once such effects become important the system cannot be studied analytically, and the complex out of equilibrium dynamics are best approached using lattice simulations. These show that perturbative inflaton decays always play a role in the final stages of reheating \cite{Khlebnikov:1996wr,Felder:2000hr,Micha:2002ey,Figueroa:2016wxr}. Consequently while preheating can have important phenomenological effects, for example in allowing for the production of heavy particles \cite{Giudice:1999fb}, it does not help realise Nnaturalness in typical oscillating inflaton models; the relative energy of the $N$ sectors is still primarily determined by the late-time perturbative decays.

Instead we study theories in which the inflaton potential has no minimum in the relevant field range, but instead becomes extremely flat, called non-oscillating (NO) models. These have previously been considered with various motivations including the possibility of realising quintessence \cite{Spokoiny:1993kt,Joyce:1997fc,Peebles:1998qn,Chung:2007vz}. The lack of inflaton oscillations means that there are no perturbative inflaton decays, and when NO models were first proposed reheating was a significant challenge. Gravitational interactions transfer some energy \cite{Ford:1986sy}, but are extremely weak and often lead to problems with overproduction of gravitons and isocurvature bounds \cite{Felder:1999pv}. Subsequently it was realised that non-perturbative processes can lead to successful reheating \cite{Felder:1998vq,Felder:1999pv}. The dynamics of these are similar to those in the oscillating inflaton case. As the inflaton evolves there can be a time when a state it is coupled to violates the adiabaticity condition, resulting in efficient particle production. Unlike the oscillating scenario, the inflaton only passes through the important region in field space once, so there is no Bose enhancement and a relatively small fraction of its energy is transferred. Despite this if the inflaton potential is sufficiently flat the remainder of its energy is kinetic and redshifts away as $1/a\left(t\right)^6$, where $a\left(t\right)$ is the scale factor of the universe. Meanwhile the inflaton decay products are assumed to be part of an interacting sector and form a relativistic plasma. The energy density in this redshifts as $1/a\left(t\right)^4$, and comes to dominate the universe.\footnote{The efficiency of reheating can be enhanced if after non-perturbative production the mass of the produced states is increased by the subsequent motion of the inflaton before they decay \cite{Felder:1999pv}. However this is not possible in Nnaturalness models since the Higgs mass cannot change by more than $\sim 100~\GeV$ after production if the hierarchy problem is to be solved.}

\subsection{Energy transfer in non-oscillating inflation models}

The energy transferred by non-perturbative reheating in NO models can be calculated following the approach developed for general theories of inflation in \cite{Kofman:1997yn}. In NO models this is simplified because there is only one period of particle production, and allowing for an arbitrary coupling between the inflaton and the decay products is straightforward. As a result we only briefly summarise the calculation, full details may be obtained from  \S VII of \cite{Kofman:1997yn}.

We study models in which the inflaton is coupled to a scalar field $\chi$. This can be expanded in terms of creation and annihilation operators $a_k^{\dagger}$ and $a_k$ where $k$ is the comoving momentum
\beq
\chi\left(x,t\right) = \frac{1}{\left(2 \pi\right)^{3/2}} \int d^3 k \left(a_k^{\dagger} \chi_k^*\left(t\right) e^{i k.x}+a_k \chi_k\left(t\right) e^{-i k.x} \right) ~.
\eeq
Following \cite{Kofman:1997yn}, we define $X_k\left( t \right) = a^{3/2}\left(t\right) \chi_k\left(t\right)$, where $a\left(t\right)$ is the scale factor of the universe, and treat the inflaton as a classical field. For simplicity we assume that the inflaton is coupled to $\chi$ through an interaction of the form 
\beq \label{eq:mint}
\mathcal{L}_{\rm int} = - \frac{1}{2}m_{\rm int}^2\left(\phi\right) \chi^2~,
\eeq
so that the inflaton gives a contribution to the mass squared parameter of $\chi$. The time evolution of an eigenfunction with momentum $k$ is then given by
\beq \label{eq:tX}
\ddot{X}_k + \left( \frac{k^2}{a^2\left(t\right)} + m_{\chi}^2 + m_{\rm int}^2 \left(\phi\left(t\right)\right) \right) X_k = 0 ~,
\eeq
where $m_{\chi}^2$ is the part of the mass squared parameter that is independent of the inflaton, and terms $\sim \left(\dot{a}/a\right)^2$ and $\ddot{a}/a$ have been dropped since these are always small in the cases of interest. 

In the adiabatic regime, which is applicable at early and late times away from the crucial moment of particle creation, Eq.~\eqref{eq:tX} is solved in terms of functions $\alpha_k\left(t\right)$ and $\beta_k\left(t\right)$ with
\beq \label{eq:Xkt}
X_k\left(t\right) = \frac{1}{\sqrt{2 \omega\left(t\right)}} \left(\alpha_k\left(t\right) e^{-i \int^t dt'~ \omega\left(t'\right)}+ \beta_k\left(t\right) e^{i \int^t dt'~ \omega\left(t'\right)} \right)~,
\eeq
where $\omega\left(t\right)= \sqrt{\frac{k^2}{a^2\left(t\right)} + m_{\chi}^2 + m^2_{\rm int}\left(\phi\left(t\right)\right)}$, and
\beq
\begin{aligned}
\dot{\alpha_k}\left(t\right) &= \frac{\dot{\omega}}{2 \omega} e^{2i \int^t dt'~ \omega\left(t'\right)} \beta_k\left(t\right) ~, \\
\dot{\beta_k}\left(t\right) &= \frac{\dot{\omega}}{2 \omega} e^{-2i \int^t dt'~ \omega\left(t'\right)} \alpha_k\left(t\right) ~.
\end{aligned}
\eeq
As discussed in \cite{Kofman:1997yn}, the occupation number of a mode is given by
\beq
\begin{aligned}
n_k &= \frac{1}{2}\left( \frac{1}{\omega_k}\left|\dot{X}_k\right|^2+ \omega_k \left|X_k\right|^2  -1\right) \\
&= \left|\beta_k\right|^2 ~.
\end{aligned}
\eeq
Production of $\chi$ states from the inflaton corresponds to the appearance of a non-zero $\beta_k$, starting from initial conditions of $\alpha_k =1$ and $\beta_k = 0$ at $t=-\infty$  (subsequently normalisation fixes $\left|\alpha_k\right|^2-\left|\beta_k\right|^2 =1$). 
Before the inflaton reaches the part of field space where particle creation happens, the modes are given by
\beq \label{eq:Xk0}
X_k^0\left(t\right) = \frac{1}{\sqrt{2 \omega_k}} e^{-i \int_{-\infty}^{t} dt'~ \omega\left(t'\right)} ~,
\eeq
while afterwards they takes the asymptotic form
\beq \label{eq:Xk1}
X_k^1\left(t\right) = \frac{\alpha_k^1}{\sqrt{2 \omega_k}} e^{-i \int_{-\infty}^{t} dt'~ \omega\left(t'\right)} +\frac{\beta_k^1}{\sqrt{2 \omega_k}} e^{i \int_{-\infty}^{t} dt'~ \omega\left(t'\right)} ~,
\eeq
where $\alpha_k^1$ and $\beta_k^1$ are constants.

Since particle production happens over a short time compared to the expansion of the universe $a\left(t\right)$ can be treated as a constant in Eqs.~\eqref{eq:tX} and \eqref{eq:Xkt}. Calculating the values of $\alpha_k^1$ and $\beta_k^1$ from Eq.~\eqref{eq:Xkt}, and therefore the number density of particles produced, is then equivalent to finding the reflection $R_k$ and transmission $D_k$ amplitudes for an incoming wave Eq.~\eqref{eq:Xk0} scattering off a potential given by $-m^2_{\rm int}\left(\phi\left(t\right)\right)$. The occupation number of $\chi$ modes after the time when production happens is given by
\beq
n_{k} = \left|\frac{R_k}{D_k}\right|^2 ~, 
\eeq
where the denominator appears so that the mode is properly normalised after production. 
Integrating over momentum modes, the total number density of produced particles is
\beq
n_{\chi}\left(t\right) = \frac{1}{2\pi^2 a\left(t\right)^3} \int dk~ k^2 n_{k} ~,
\eeq
and their energy density immediately after reheating is
\beq
\rho_{\chi}\left(t\right) = \frac{1}{2\pi^2} \int dk ~ k^2 \sqrt{k^2+ m_{\chi}^2 + m^2_{\rm int}\left(\phi\left(t\right)\right)} n_{k} ~.
\eeq

\subsection{Example inflaton couplings}

A particularly simple case is if production is efficient only for a short time around the moment when the decay product's mass takes its minimum value (when the inflaton is at $\phi_0$). In this part of field space the time dependence of $\chi$'s mass squared parameter can be expanded in Eq.~\eqref{eq:tX} as
\beq \label{eq:quadc}
\begin{aligned}
m_{\chi}^2 + m_{\rm int}^2\left(\phi\right)  &\simeq   m_{\chi}^2  + m_{\rm int}^2\left(\phi_0\right)  + \frac{1}{2}\left(t -t_0 \right)^2  \dot{\phi}^2  \frac{\partial^2 m^2_{\rm int}\left(\phi_0\right)}{\left(\partial \phi\right)^2} \\
&\equiv  m_0^2 + \frac{1}{2} g^2 \left(t -t_0 \right)^2 \dot{\phi}^2  ~,
\end{aligned}
\eeq
where $m_{0}^2$ is the total mass of $\chi$ at $\phi=\phi_0$, $g^2$ is the effective quartic coupling between the inflaton and $\chi$ at this point, and $\dot{\phi}$ is the inflaton's approximately constant speed. The occupation numbers of the produced particles are then given by the reflection and transmission coefficients for scattering off a negative quadratic potential. These can be straightforwardly calculated with standard techniques \cite{landau1977quantum}, leading to
\beq \label{eq:nk}
n_{k} = \exp\left(- \frac{\pi \left(k^2+ m_0^2\right)}{g \left|\dot{\phi}\right|}\right) ~,
\eeq
and therefore
\beq \label{eq:nquartic}
n_{\chi}\left(t_0\right) = \frac{1}{8 \pi^3} \left(g \left|\dot{\phi} \right|\right)^{3/2} \exp\left(- \frac{\pi m_0^2}{g \left|\dot{\phi}\right|}\right) ~.
\eeq
As anticipated, the number density of produced states is exponentially suppressed when the adiabaticity condition $\dot{m}/m_{0}^2 = g \dot{\phi}/ m_0^2 \lesssim 1$ is satisfied.

A decrease and then increase of the decays product's mass is not required for particle production. For example, with an interaction 
\beq \label{eq:expint}
\mathcal{L}_{\rm int}= - \frac{1}{2} g^2 \Lambda^2 \exp\left(-\phi/\Lambda\right) \chi^2 ~,
\eeq
and the inflaton starting at $- \infty$ and moving to $+ \infty$, there is no finite point in field space at which the mass of $\chi$ is minimised, however reheating still occurs. Assuming the inflaton has a constant speed,
  in the adiabatic limit the occupation number of the produced modes is given by
\beq \label{Eq:nnk}
n_{k} \simeq \frac{4 \sqrt{k^2 + m_{\chi}^2}}{g \Lambda} \exp\left(-8 \pi  \frac{\Lambda \sqrt{k^2 + m_{\chi}^2}}{\dot{\phi}}  \right) ~. 
\eeq
The maximum value of $\dot{m}_{\rm int}/\left(m_{\rm tot}^2+k^2\right)$ during the inflaton's evolution (where ${m}_{\rm int}^2$ is the inflaton dependent part of $\chi$'s mass squared and $m_{\rm tot}^2$ is its total mass squared) 
is $ \sim \dot{\phi}/\left(4 \Lambda \sqrt{m_{\chi}^2+k^2} \right)$, so the suppression in Eq.~\eqref{Eq:nnk} has the expected parametric dependence. Away from the adiabatic limit the number density of particles produced is comparable to that from a quartic interaction with the same value of $g$, given by Eq.\eqref{eq:nquartic}.\footnote{For the coupling Eq.~\eqref{eq:expint} the general expression for the occupation number after reheating is a combination of special functions, and can be obtained from the reflection and transmission amplitudes given in e.g. \cite{2003quant.ph..9176A}.}

The production rate is not always exponentially suppressed in the limit $\dot{m}/m^2 \ll 1$. For example, an inflaton coupling
\beq
\mathcal{L}_{\rm int} = \begin{cases}
		    -\frac{1}{2} g^2 \phi^2 \chi^2 , & \phi < 0 \\
		    0,& \phi \geq 0 
                    \end{cases} ~,
\eeq
corresponding to reflection off a negative quadratic potential in the region $t<0$, can be calculated analytically, again assuming a constant inflaton speed (expressions for the coefficients are given in e.g. \cite{0305-4470-30-14-031,0305-4470-31-13-014}). In the adiabatic limit $ \left(m_{\chi}^2 + k^2 \right)/ \left(g \dot{\phi} \right) \gg 1$ the occupation number of the produced states is
\beq
n_k = \frac{1}{4} \left(\frac{g \dot{\phi}}{m_{\chi}^2 + k^2}\right)^2 ~,
\eeq
i.e. power law suppressed in the parameter $\dot{m}_{\rm int}/\left(m_{\chi}^2 + k^2\right)$. Investigation of example potentials suggests that a power law rather than exponential suppression happens when the first time derivative of the mass of $\chi$ is everywhere small but a higher derivative becomes large at some point (or the mass is not a smooth function of time). In practice the inflaton couplings we consider for Nnaturalness models lead to either exponential suppression or a suppression by a high power of the parameter $\dot{m}/m^2$ allowing for viable phenomenology. 

While the rate of particle production can also be found analytically for more complex potentials, in the present work we instead calculate the scattering and reflection coefficients numerically, solving Eq.~\eqref{eq:Xkt} subject to the boundary conditions Eqs.~\eqref{eq:Xk0} and \eqref{eq:Xk1}. This is straightforward, and the results obtained can be understood from simple physical arguments (meanwhile away from the adiabatic limit analytic results are typically combinations of special functions).

\section{Nnaturalness and non-oscillatory potentials} \label{sec:3}

We now study how non-perturbative reheating in non-oscillatory theories can be used to build models of Nnaturalness. As mentioned, the core idea of Nnaturalness is to consider a theory containing $N$ copies of the visible sector with varying Higgs mass squared parameters. For simplicity we follow \cite{Arkani-Hamed:2016rle} and assume that these are otherwise identical, however in general the other parameters of the sectors, or even their field content or gauge groups,  could also change without destroying the mechanism. The  distribution of the Higgs mass squared parameters in the different sectors is taken to be flat up to a cutoff $\Lambda^2$.\footnote{This is usually assumed when arguing for the existence of a hierarchy problem in the first place.} Then in the sector with the lightest Higgs this typically has mass of order $\sim \Lambda/\sqrt{N} $, while in the sector with the next lightest it is expected to be roughly in the range $\sqrt{2} \Lambda/ \sqrt{N} \div 2 \Lambda/ \sqrt{N} $.

Constraints on the number of extra relativistic degrees of freedom from CMB and BBN measurements immediately rule out the possibility that all sectors have comparable energy densities at late times. Additionally, this scenario would not directly solve the hierarchy problem without anthropic reasoning, the question would simply be shifted to why we reside in the particular sector that has a small EW VEV. Instead dynamics are needed that result in only a sector with a light Higgs and non-zero VEV being reheated significantly after inflation, while the other sectors remain cold and presumably unsuitable for life.

Non-perturbative reheating can allow Nnaturalness models if the $N$ sectors are reheated through a coupling between their Higgses and the inflaton. As an example, suppose the Higgses have couplings to the inflaton that can be approximated as quartic interactions, Eq.~\eqref{eq:quadc}, in the relevant part of field space. The non-perturbative energy transfer to a sector will be highly suppressed if $g \left|\dot{\phi}\right| \lesssim m_0^2$, where $m_0^2$ is the mass squared parameter of its Higgs at the moment of production. Since it is technically natural for $g$ to be parametrically small, this is sensitive to the scale $ g \left|\dot{\phi}\right|$, which can be much below any others in the theory without fine tuning. Consequently the sector containing the lightest Higgs could be dominantly reheated and identified as the one we live in, with exponentially less energy transferred to the others.

In the present work we study a toy version of the SM without including the possible effects of gauge bosons or fermions. Both gauge bosons \cite{GarciaBellido:1999sv,Rajantie:2000fd,GarciaBellido:2003wd,DiazGil:2007dy,Adshead:2012kp,Deskins:2013lfx} and fermions \cite{Baacke:1998di,Greene:1998nh,Giudice:1999fb,Greene:2000ew} can be produced non-perturbatively through dynamics similar to scalars. In the SM the EW gauge boson and the fermion masses depend on the Higgs VEV, and non-perturbative production of these is possible in sectors in which the Higgs VEV changes during the inflaton's evolution.  

The production of fermion modes that get a mass from a Yukawa coupling $y_f$ with a Higgs is likely to be exponentially suppressed in the limit
\beq
\frac{\dot{m}_f}{m_f^2 + k^2} \sim \frac{y_f \dot{\left<h\right>}}{y_f^2 \left<h\right>^2+ k^2} \ll 1 ~,
\eeq
similarly to Eq.~\eqref{eq:nk}. 
For a fermion with a small Yukawa coupling, production is possible in all sectors satisfying $\dot{\left<h\right>} \gtrsim y_f \left<h\right>^2 $. However, the total energy transfer is small since only modes with momentum $k^2 \lesssim y_f \dot{\left<h\right>}$ are created. Since $\dot{\left<h\right>}$ is suppressed by a coupling constant in the models of interest, only fermions with tiny $y_f$ are in this regime, and the energy transferred $\sim k^4 \lesssim y_f^2 \dot{\left<h\right>}^2$ is much less than that directly to a light Higgs. Similarly to the production of Higgses the scale $\dot{\left<h\right>}$ is involved, in contrast to models with perturbative inflaton decays. Non-perturbative energy transfer to fermions with large Yukawa couplings might be important, in particular to the top quark. Production will only be efficient to sectors in which $\dot{\left<h\right>} \sim \left<h\right>^2$ at some point during the inflaton's evolution, which are exactly those with a small non-zero Higgs VEV. Similar arguments hold for non-perturbative production of the gauge bosons through the dependence of their masses on the Higgs VEV. 

There is also an inflaton interaction with the ${\rm SU}\left(2\right)$ gauge bosons radiatively generated through the Higgses, suppressed by $\sim 1/m_h^2$ \cite{Arkani-Hamed:2016rle}. This is present both in sectors with zero and non-zero Higgs VEVs, and once EW symmetry is broken either by a Higgs VEV or QCD includes a coupling to the photon. While it can lead to non-perturbative production, this coupling is suppressed by a loop factor compared to the inflaton's interactions with the Higgses. Therefore even for light gauge bosons only low momentum modes can be produced, and we anticipate that the energy transfer is relatively inefficient especially to sectors with large $m_h$.

As a result, if it turns out to be significant, the possible extra energy transfer due to including fermions and gauge bosons will plausibly favour the correct sectors for an implementation of Nnaturalness. While a full analysis is extremely important for a complete model, we leave this to future work.

\subsection{Constraints on the inflation couplings and potential}

There are immediate challenges to constructing Nnaturalness models using non-perturbative reheating. As discussed the inflaton potential must be of a non-oscillating form (of course viable models are still possible if this is not the case, but the exponential suppression of preheating would play no role). It must also be extremely flat in the part of field space probed after reheating, so that the inflaton energy is dominantly kinetic and redshifts fast, and the reheated sector comes to dominate the energy density of the universe. The Hubble parameter immediately after reheating is $\sim \dot{\phi}/M_{\rm Pl}$ (where $M_{\rm Pl}$ is the Planck mass) and, with the required flat potential, the subsequent evolution of the inflaton is
\beq
\phi\left(a\left(t\right)\right) \sim M_{\rm Pl} \log \left(\frac{a\left(t\right)}{a_0} \right)~,
\eeq
where $a_0$ is the scale factor of the universe at reheating (backreaction from the produced states is always negligible in phenomenologically interesting models). The Planck scale field ranges involved are potentially worrying for obtaining a viable UV completion, and the inflaton couplings to the $N$ sectors must be such that this flatness is technically natural. A further constraint is that in moving to such large field values the inflaton cannot change the Higgs mass squared parameter of the reheated sector by more than $\sim \left(100~\GeV\right)^2$, otherwise the sector with the lightest Higgs at the moment of particle production will not have a light Higgs at late times.

These issues can be evaded by assuming the inflaton's couplings to the $N$ sectors vanish at large field values. As an example we consider interactions of the form
\beq \label{eq:infcou}
\mathcal{L}_{\rm int} = - g^2 \Lambda^2 \left|h\right|^2 e^{- g \phi/\Lambda}~,
\eeq
with inflation ending and reheating beginning at $\phi = 0$ with $\dot{\phi}>0$. Eq.~\eqref{eq:infcou} leads to an inflaton dependent component of the Higgs mass squared $m_{\rm int}^2$ as in Eq.~\eqref{eq:mint}. 

When studying the energy transfer we will assume that the inflaton speed is comparable to the cutoff of the theory $\dot{\phi} \sim \Lambda^2$. This is consistent with Eq.~\eqref{eq:mint} without fine tuning provided $N g^2 \lesssim 1$, which will be the case due to other phenomenological constraints.\footnote{More generally if the inflaton couplings to the different sectors were different enough that they gave random sign contributions to the inflaton potential, the sum of these would have a dependence $\sim \sqrt{N}$ allowing smaller values of $\dot{\phi}$.} Eq.~\eqref{eq:infcou} leads to an inflaton potential near the origin $V \supset g^2 \Lambda^2 \phi^2 $, suppressed by $g^2$. However this is not an essential feature, and the inflaton potential can include a mass term $\sim \Lambda^2 \phi^2$ close to $\phi=0$ without altering our results provided its speed remains $\dot{\phi} \lesssim \Lambda^2$. Once the inflaton has moved to field values $\phi \gg \Lambda/g$, its couplings to the Standard Model like sectors are strongly suppressed and do not ruin the assumed flatness of its potential. The results we obtain are not dependent on Eq.~\eqref{eq:infcou} remaining valid once $\phi \gg \Lambda/g$, provided that the couplings vanish sufficiently fast.

While unfortunately we do not have explicit examples of UV models that lead to the assumed interactions, they could plausibly come from a model in which the inflaton has a shift symmetry broken by strong coupling dynamics when it is close to the origin. Similarly, we do not attempt to construct a full model for the inflaton potential. Instead we simply assume that a phenomenologically viable period of inflation happens for $\phi <0$, and that at the end of this the inflaton starts moving towards $\phi = \infty$, with a potential that vanishes at large field values. Although by no means a viable candidate for an actual model (e.g. due to its Planck suppressed couplings) a hint that the required features might be possible comes from the string theory dilaton. A dilaton potential at finite field values can be generated by hidden sector gaugino condensation. However, since the dilaton VEV sets the gauge couplings, at large field values its potential vanishes leading to a run away to infinity \cite{Dine:1985rz}. Flat directions are also common in many models of supersymmetry breaking \cite{Shirman:1996jx}.

\subsection{Energy transferred}

The success of the present implementation of Nnaturalness rests on how the energy transferred to a sector depends on the inflaton independent part $m_h^2$ of its Higgs squared mass parameter, which given Eq.~\eqref{eq:infcou} is also the late-time total Higgs mass squared parameter once $\phi\rightarrow \infty$. To calculate this we numerically evaluate the reflection and transmission coefficients for scattering off the corresponding potential, as described in Section~\ref{sec:2}. The inflaton decouples from the Higgses after evolving a distance $\sim \Lambda/g$ and during this time the Higgs mass squared parameters change by less than $\sim \left(100~\GeV\right)^2$, so there is no ``fattening'' of the produced states unlike the models \cite{Felder:1999pv,Kofman:2004yc}. Consequently only a small fraction of the inflaton's energy is transferred and there there is no backreaction on its motion (this is not altered by the large number of sectors since particle production is exponentially suppressed in the vast majority of these).

A case of particular interest is when the total mass squared parameter of the Higgs in a sector changes from positive to negative during the inflaton's evolution. While the adiabaticity condition is satisfied, the Higgs tracks the minimum of its potential closely, but around the moment when the total mass squared parameter changes sign the evolution is non-adiabatic. However for our current purposes it is sufficiently accurate to assume that the Higgs tracks the instantaneous minimum of its potential exactly throughout. As an additional simplification we neglect the mixing between that Higgs and the inflaton that arises from Eq.~\eqref{eq:infcou} when $\left<h\right> \neq 0$ (such mixing terms have been found to be important in some other theories with preheating, e.g. \cite{BasteroGil:1999fz}). Consequently the results of Section~\ref{sec:2} are assumed to be directly applicable, with the inflaton coupling to a complex scalar when $\left<h\right> = 0$, and the Higgs mode when $\left<h\right> \neq 0$ (with the corresponding physical masses). The mixing terms in Eq.~\eqref{eq:infcou} are $\sim g^3 \Lambda \left<h\right>$, and in the models we consider this is suppressed relative to the Higgs mass squared even in the sector in which this is smallest (which will turn out to be $\sim g^2 \Lambda^2$). Consequently ignoring mixing is a reasonable first assumption, although we stress that more detailed future study is required.

If we take the coupling Eq.~\eqref{eq:infcou}, and assume the inflaton is fixed at $\phi=0$ until $t=0$ and then has a constant speed, $\dot{m}_{\rm int}$ is not continuous. While this still leads to phenomenologically viable models it does result in a power law rather than exponential suppression of the energy transfer to sectors with a heavy Higgs. Instead we fix the time dependence of the Higgs masses to be
\beq \label{eq:c2}
\mathcal{L}_{\rm int} = - g^2 \Lambda^2 \frac{\left|h\right|^2}{\exp\left(\frac{\sqrt{8} g \dot{\phi}t}{\Lambda}\right)+1} ~,
\eeq
for $-\infty <t < \infty$ with $\dot{\phi}$ constant. This corresponds to   giving the inflaton speed a time dependence in Eq.~\eqref{eq:infcou} (the coefficient $\sqrt{8}$ is chosen so that $\dot{m}_{\rm int}$ is approximately the same as Eq.~\eqref{eq:infcou} close to $\phi = 0$). In the limit of large $m_h^2$ or a high momentum mode, the maximum value of $\dot{m}_{\rm int}/\left(\left|m_h^2 + m_{\rm int}^2 \right|+k^2\right)$ as the inflaton evolves is parametrically $\sim \sqrt{g^2 \dot{\phi}/\left(\left|m_h^2\right|+ k^2\right)}$ and away from this limit the behaviour is more complicated.

\begin{figure}
	\begin{center}
		\includegraphics[width=.485\textwidth]{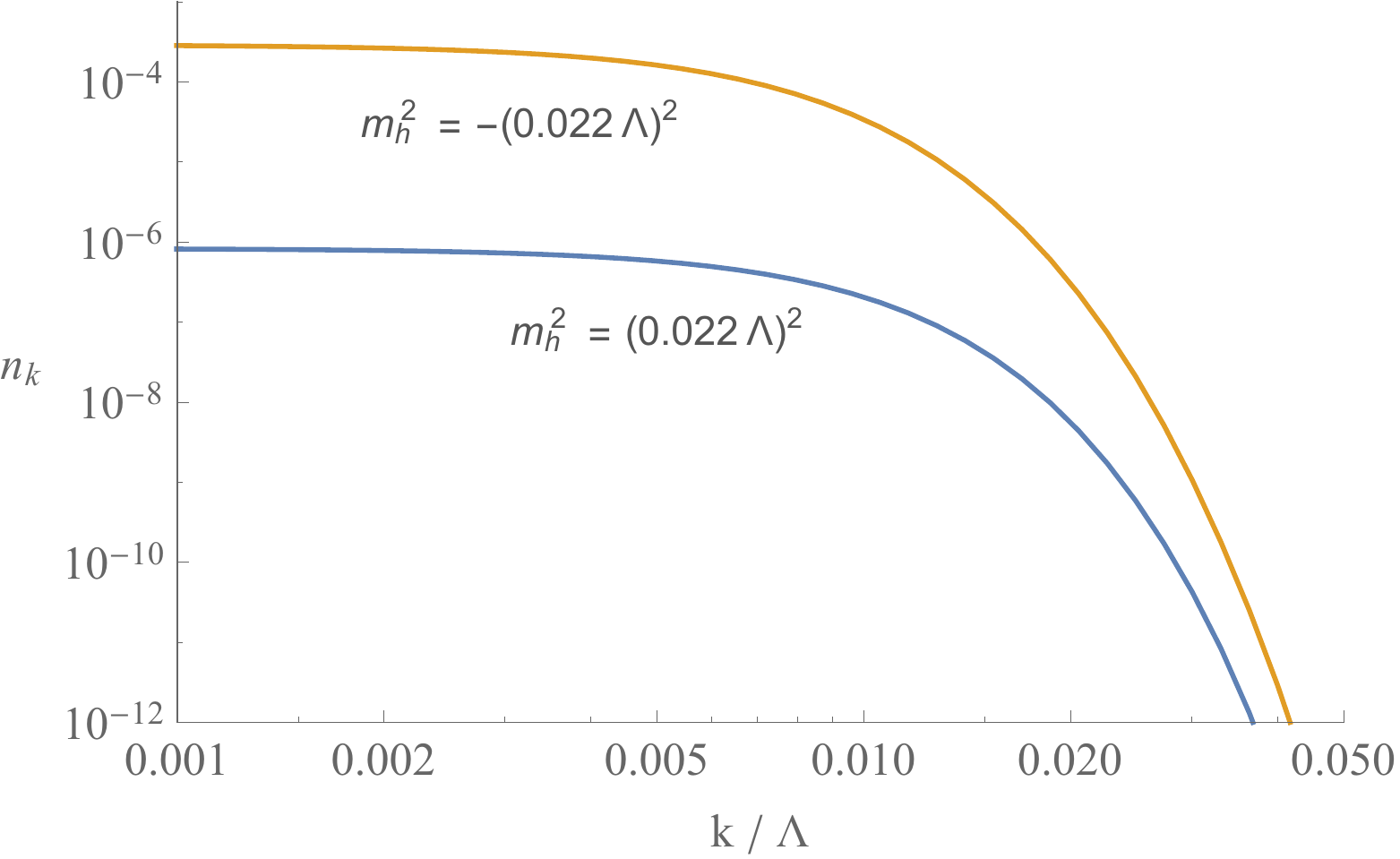}
		~
		\includegraphics[width=.485\textwidth]{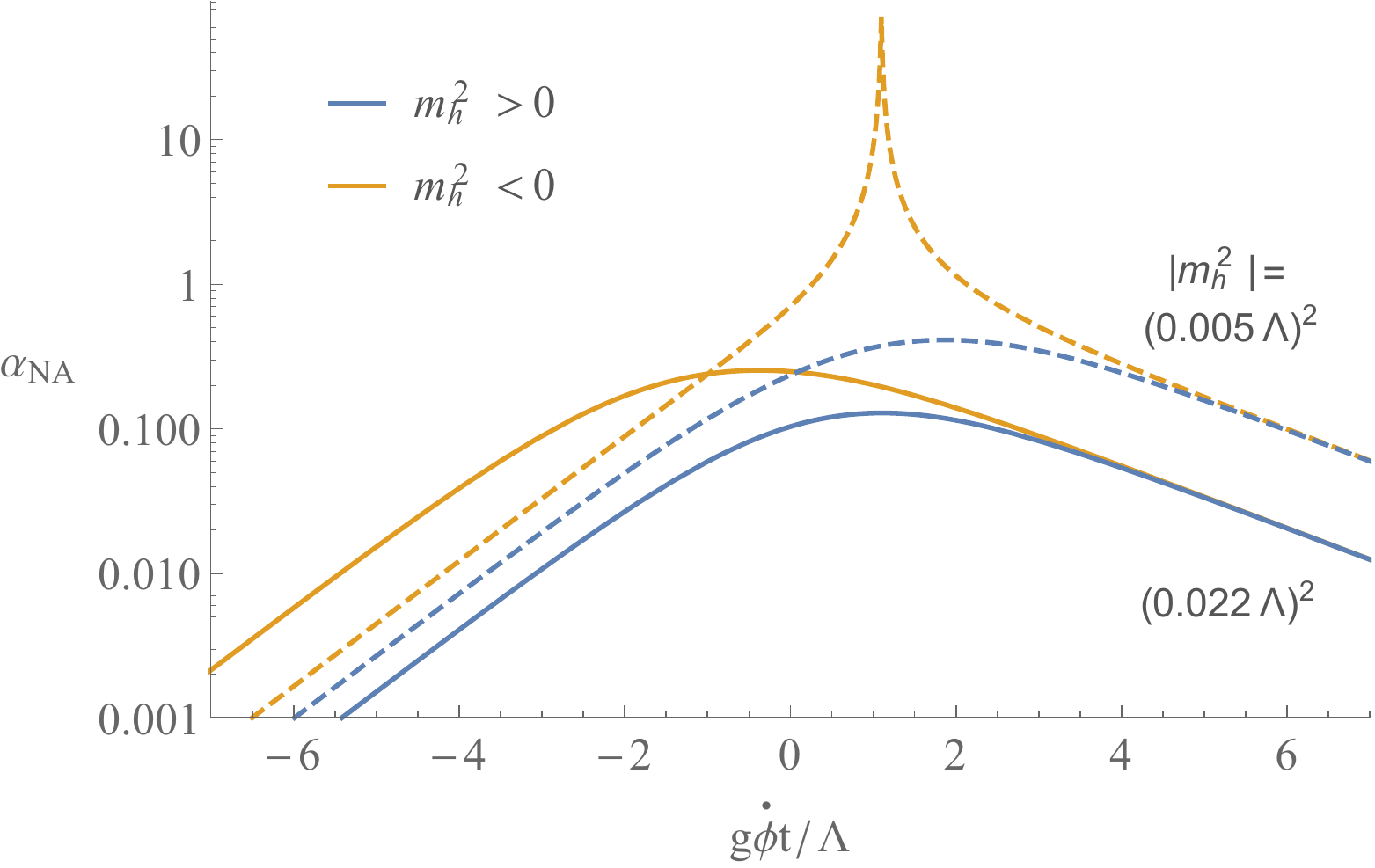}
	\end{center}
	\caption{{\bf \emph{Left:}} The occupation number $n_k$ of Higgs modes with momentum $k$ produced during reheating, for sectors in which the inflaton independent components of the Higgs mass squared parameters are $m_h^2 = \left(0.022\Lambda\right)^2$ and $m_h^2 = -\left(0.022\Lambda\right)^2$. The inflaton couplings are given by Eq.~\eqref{eq:c2}, with $g=0.02$ and a constant speed $\dot{\phi} = \Lambda^2 / \sqrt{8}$. A partial cancellation in the total Higgs mass while $\dot{\phi} t \lesssim 0$ in the negative Higgs mass squared case leads to more efficient production of low momentum modes.  {\bf \emph{Right:}} The value of the non-adiabaticity measure $\alpha_{\rm NA}=\dot{m}_{\rm int}/\left(\left|m_h^2 + m_{\rm int}^2 \right|+k^2\right)$ as a function of time, where $m_{\rm int}^2$ is the inflaton dependent component of the Higgs mass squared parameter, for a mode with momentum $k = 0.001 \Lambda$. The results are shown for sectors with different Higgs mass squared parameters, and the inflaton couplings are as in the left panel. For $m_h^2 = - \left(0.022 \Lambda\right)^2$ the physical Higgs mass is small over a large time range $\dot{\phi} t \lesssim 0$, and production is  more efficient than for  $m_h^2 =  \left(0.022 \Lambda\right)^2$. For $m_h^2 = - \left(0.005 \Lambda\right)^2$ there is a moment when the physical Higgs mass vanishes, leading to more energy transfer than for $m_h^2 = \left(0.005 \Lambda\right)^2$.
	}
	\label{fig:1}
\end{figure}

In Figure~\ref{fig:1} left, we plot the occupation number of Higgs modes produced by reheating as a function of their momentum $k$, for sectors in which $m_h^2 = \left(0.022\Lambda\right)^2$ and $m_h^2 = -\left(0.022\Lambda\right)^2$ with $g=0.02$ and $\dot{\phi} = \Lambda^2/ \sqrt{8}$. In Figure~\ref{fig:1} right, the non-adiabaticity measure $\dot{m}_{\rm int}/\left(\left|m_h^2 + m_{\rm int}^2 \right|+k^2\right)$ is shown as a function of time for the same values of $m_h^2$ and also for  $m_h^2 = \pm \left(0.005 \Lambda\right)^2$.

In sectors with $\left|m_h^2\right| > g^2 \Lambda^2$, adiabaticity is maximally violated fairly close to $\dot{\phi} t \simeq 0$ since $\dot{m}_{\rm int}$ is largest at this moment. The numerical results in Figure~\ref{fig:1} left for $m_h^2 = \left(0.022 \Lambda\right)^2$ are accurately fit by 
\beq \label{eq:nkft}
n_k \simeq 80 \exp\left(-14\sqrt{\left(k^2+ \left|m_0^2\right|\right)/\left(g \dot{\phi} \right)} \right) ~,
\eeq
where $m_0^2 = \left( 0.022 \Lambda\right)^2 + \left(0.010 \Lambda\right)^2 $, and in fitting this we have assumed the parametric form and the value of $m_0$. The same expression with $m_0^2 = 2\left( \left(0.022 \Lambda\right)^2 -  \left(0.010 \Lambda\right)^2 \right)$ fits the high momentum modes for the $m_h^2 <0$ case well (the factor $2$ is due to the different physical mass when the Higgs has a VEV). Low momentum modes are enhanced by an extra factor of $5$, beyond that from this modification of $m_0$. The extra energy transfer happens due to production at times $\dot{\phi} t < 0$, when $\dot{m}$ is smaller but there is a greater cancellation between the bare mass term and the inflaton induced one. Similarly for $m_h^2 > \Lambda^2$ production dominantly occurs at $\dot{\phi} t > 0$, when the total Higgs mass is smaller. This is reflected in the non-adiabaticity parameter, which is shown in Figure~\ref{fig:1} right as a function of time. Because production does not occur exactly at $\dot{\phi} t =0$, Eq.~\eqref{eq:nkft} is not accurate for arbitrary $m_h^2$ unless the values of the coefficients are also varied as a function of $m_h^2$ .\footnote{This is also likely to account for the $> 1$ prefactor in Eq.~\eqref{eq:nkft}, although it is not physically forbidden.}

In a sector with $-g^2 \Lambda^2 <  m_h^2 < 0$ there is a moment during the inflaton's evolution when the total mass of its Higgs vanishes. The strong violation of adiabaticity around this time, visible in Figure~\ref{fig:1} right, significantly increases the energy transfer compared to a sector with a positive Higgs mass squared parameter of the same magnitude. 
  Reducing $\left|m_h^2\right|$ the cancellation happens later, when $\dot{m}_{\rm int}$ is smaller, so particle production is less efficient.\footnote{In the $-g^2 \Lambda^2 <  m_h^2 < 0$ case the high momentum modes are actually only power law rather than exponentially suppressed. This is due to $\dot{m}_{\rm int}$ not being smooth when the total effective mass squared parameter passes through zero. However, these modes are not important for the total energy transferred (since the power law is stronger than $\sim k^{-4}$), and modifying the evolution of the mass so that $\dot{m}_{\rm int}$ is smooth around the critical point, and the high modes are exponentially suppressed, does not change the energy transferred significantly.}

\begin{figure}
	\begin{center}
		\includegraphics[width=.6\textwidth]{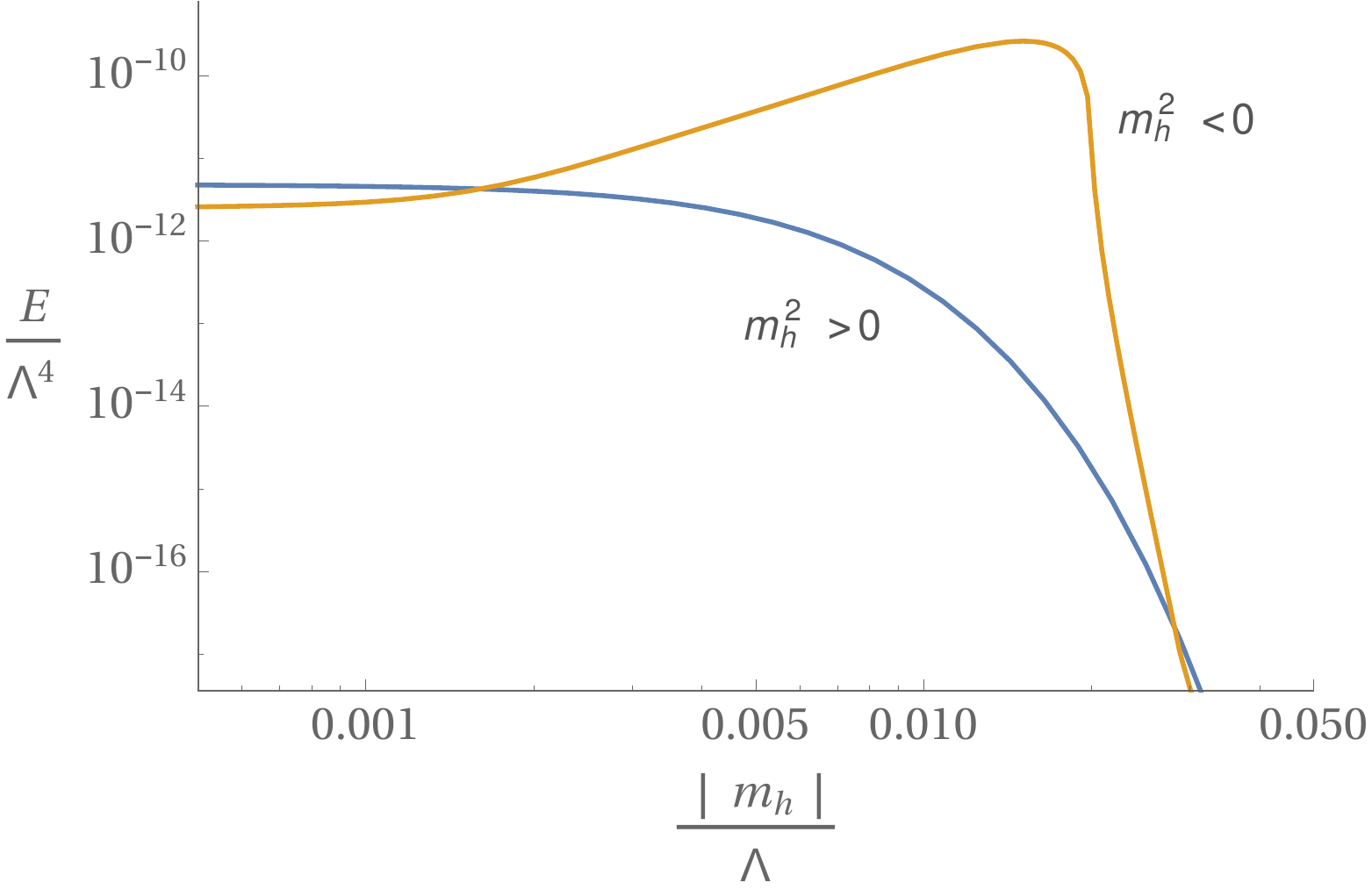}
	\end{center}
	\caption{
	The energy $E$ transferred to a sector during reheating as a function of the inflaton independent component of its Higgs mass squared $m_h^2$ relative to the UV cutoff of the theory $\Lambda$. The inflaton couplings are given by Eq.~\eqref{eq:c2} with $g= 0.02$ and $\dot{\phi} =  \Lambda^2 /\sqrt{8}$, and at late times the inflaton decouples and the total Higgs mass squared parameter is $m_h^2$. Above $\left|m_h\right| \sim 0.02 \Lambda$ the evolution is adiabatic and the energy transfer is exponentially suppressed. The inflaton couplings are such that if $-g^2 \Lambda^2 <m_h^2<0$ there is a moment when the physical Higgs mass vanishes, and significantly more energy is transferred than if $m_h^2>0$.
	}
	\label{fig:2}
\end{figure}

The total energy transferred to a sector as a function of the late-time mass of its Higgs is shown in Figure~\ref{fig:2}, for the same inflaton couplings and parameters as Figure~\ref{fig:1}. As anticipated this is exponentially suppressed at large Higgs masses, and for a positive Higgs mass squared parameter the energy transferred simply increases as $m_h^2$ decreases, becoming saturated once $g^2 \dot{\phi} \gg m_h^2$ (similarly to the simple case of the quartic coupling in Eq.~\eqref{eq:quadc}). Meanwhile, there is a substantial boost when the Higgs mass squared parameter is negative, compared to a sector with a positive Higgs mass squared parameter of the same magnitude, with reheating most efficient for $m_h^2 \simeq -g^2\Lambda^2$.

In Figure~\ref{fig:2} the interaction Eq.~\eqref{eq:infcou} is such that the peak in energy transfer corresponding to a cancellation between $m_h^2$ and the inflaton contribution to the Higgs mass squared $\simeq g^2 \Lambda^2$ coincides with the Higgs mass at which the adiabatic suppression becomes large, $\left|m_h^2\right| \simeq g \dot{\phi}$. However this assumption is not required and can change with interactions different from Eq.~\eqref{eq:infcou}, or other values of $\dot{\phi}$. If the adiabatic suppression happens at $\left|m_h^2\right|$ larger than when there is a cancellation, the relative enhancement of the peak is reduced, since the energy transfer to sectors with $m_h^2$ the same magnitude but positive is less suppressed. In the converse case the prominence of the peak is increased.

While we have calculated the energy transfer in a simple model, similar results are obtained from more complex couplings provided the conditions discussed in Section~\ref{sec:2} are satisfied. For example, an extra coupling of the form $\mathcal{L}_{\rm int} =g^4 \phi^2 \left|h\right|^2 \exp \left(-g \phi/\Lambda \right)$ could be included in Eq.~\eqref{eq:infcou} (mapping onto an extra term in Eq.~\eqref{eq:c2} after allowing for an inflaton time dependence).\footnote{A term $g^2 \phi^2 \left|h\right|^2 \exp\left(-g \phi/\Lambda \right)$ not suppressed by $g^2$ would change the Higgs masses too much, but is not radiatively generated starting from Eq.~\eqref{eq:infcou}.} The energy transferred to sectors with a Higgs mass larger than some scale is generically strongly suppressed, and there is typically the possibility of enhanced reheating to sectors in which the total Higgs mass vanishes at some moment during the inflaton's evolution.

\section{Phenomenological constraints}\label{sec:4}

Having seen that non-perturbative processes can dominantly reheat sectors with a small negative Higgs mass squared parameter, we turn to the phenomenological constraints that must be satisfied for a viable realisation of Nnaturalness. These are similar to those in the original model \cite{Arkani-Hamed:2016rle}, although the decoupling of the inflaton at late times means that some are absent in the present implementation (on one hand avoiding potential constraints can simplify model building, but it has the downside of reducing the observable signals).

The Hubble parameter during inflation must be below the EW scale, so that the Higgs masses are sufficiently close to their zero temperature values that the correct sector is reheated. As a result the Higgses also remain close to the minimum of their potentials during inflation. This will be easily satisfied once the other constraints are imposed if the Hubble scale during inflation is $\sim \dot{\phi}^2/M_{\rm Pl}$. Low scale inflation is cosmologically viable, and could have beneficial features for model building such as allowing the initial inflaton value to be set by a finite temperature potential \cite{German:2001tz} (obtaining sufficiently homogeneous initial conditions may be challenging, and might require some earlier inflationary epoch however). Since the Hubble parameter is small, gravitational production never transfers any significant energy during reheating.

BBN \cite{Cooke:2013cba} and CMB limits \cite{Ade:2015xua} on the number of extra relativistic degrees of freedom are potentially important (and are also a significant constraint in the original model \cite{Arkani-Hamed:2016rle}). These arise due to energy transferred during reheating to the other sectors, beyond the visible one. The bounds are normally phrased in terms of an effective number of additional neutrinos $\Delta N_{\rm eff}$, and to be consistent with observations the total contribution from all the hidden sectors must be approximately $\Delta N_{\rm eff} \lesssim 0.6$. More detailed discussion including the difference between free streaming and perfect fluid radiation can be found in e.g. \cite{Bell:2005dr,Chacko:2015noa}, but for our present purposes this simple expression is sufficient. Assuming that the extra sectors are all identical copies of the visible sector apart from their differing Higgs masses, the total contribution to $\Delta N_{\rm eff}$ is approximately
\beq
\Delta N_{\rm eff} \simeq \sum_{i} 6.14 \frac{\rho_{i}}{\rho_{\rm vis}} ~,
\eeq
where $i$ runs over all the sectors other than the visible one, and $\rho_i$ and $\rho_{\rm vis}$ are the energy densities of a hidden sector and the visible sector respectively. Therefore the next most strongly reheated sector must get at least an order of magnitude less energy than the visible sector during reheating.

\begin{figure}
	\begin{center}
		\includegraphics[width=.65\textwidth]{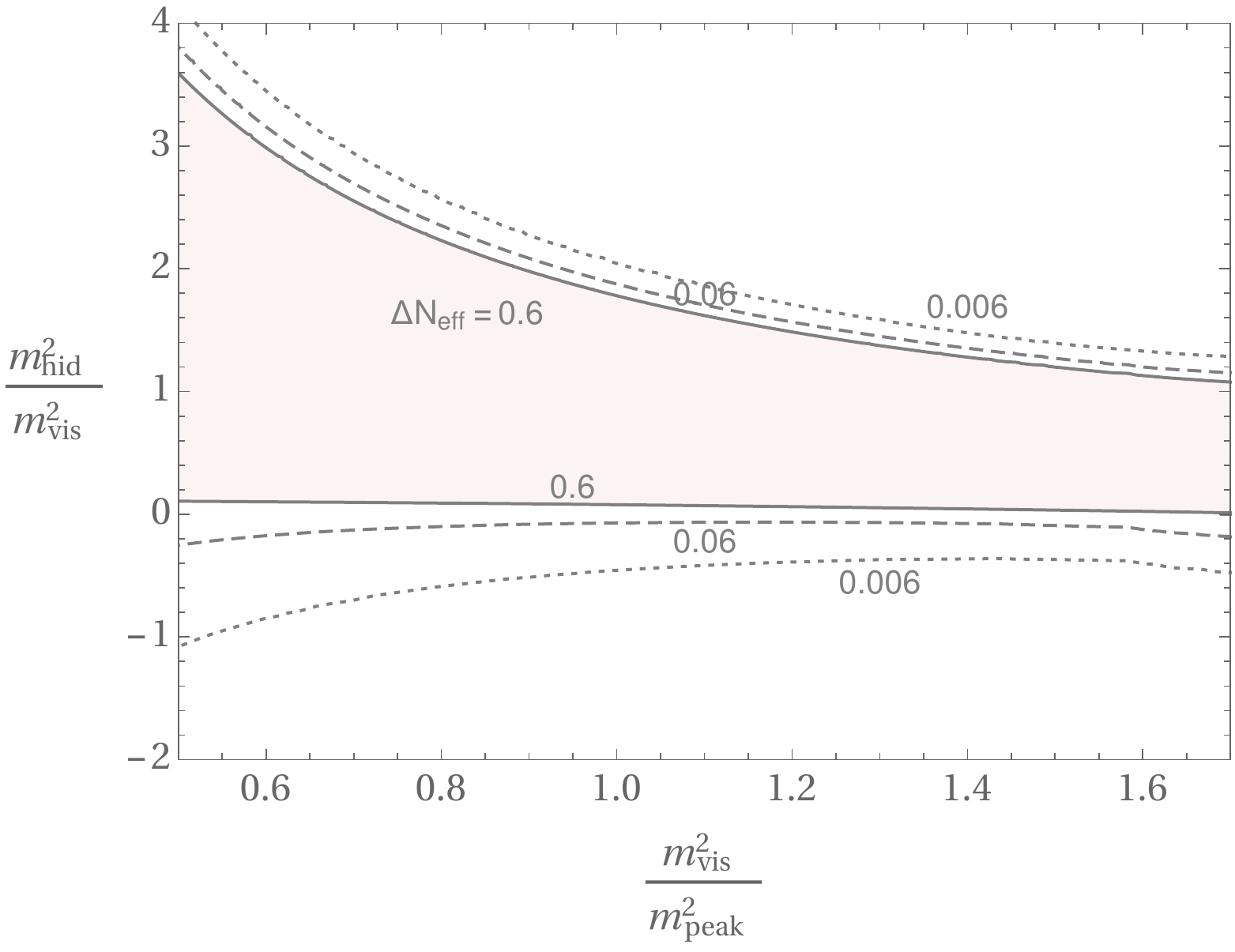}
	\end{center}
	\caption{Contours of the contribution to $\Delta N_{\rm eff}$ from a hidden sector with a (late-time) Higgs mass squared parameter $m_{\rm hid}^2$, as the visible sector Higgs mass squared parameter $m_{\rm vis}^2$ varies. The inflaton couplings are given by Eq.~\eqref{eq:c2} with $g=0.02$ and $\dot{\phi}=\Lambda^2/\sqrt{8}$, and the visible Higgs mass squared is normalised relative to the value that leads to the maximum energy transfer, $m_{\rm peak}^2 \simeq -\left(0.015 \Lambda\right)^2$ (these parameters are the same as in Figure~\ref{fig:2}). If $\left|m_h^2\right| \gtrsim  1.7 \left|m_{\rm peak}^2\right|$ or  $\left|m_h^2\right| \lesssim  0.5 \left|m_{\rm peak}^2\right|$ the energy transfer to the visible sector is suppressed, and radiation domination will not begin before BBN if the UV cutoff is $\Lambda \simeq 10~\TeV$.
	}
	\label{fig:3}
\end{figure}

In Figure~\ref{fig:3}, we plot contours of the contribution to $\Delta N_{\rm eff}$ from an additional sector as a function of its Higgs mass squared parameter $m_{\rm hid}^2$ and the visible sector Higgs mass squared parameter $m_{\rm vis}^2$. The visible sector Higgs mass squared is normalised relative to the value that maximises the energy transfer $m_{\rm peak}^2 \simeq -g^2 \Lambda^2$, and the inflaton couplings are given by Eq.~\eqref{eq:c2} with the same values of $g$ and $\dot{\phi}$ as in Figure~\ref{fig:2}. A sector with a positive Higgs mass squared parameter of the same magnitude as that of the visible sector can be easily accommodated within current limits. Meanwhile sectors with $m_h^2<0$ can lead to important constraints, but due to the exponential suppression typically only the one with the next smallest Higgs mass gives a significant contribution. This can be seen from the fast drop off in $\Delta N_{\rm eff}$ as a function of $m_{\rm hid}^2$ in Figure~\ref{fig:3} as well as in Figure~\ref{fig:2}.

The natural range for the visible sector Higgs mass squared parameter without tuning is approximately a factor of $2$ around $m_{\rm peak}^2$. Similarly the sector that dominantly contributes to $\Delta N_{\rm eff}$, is expected to have a Higgs mass squared parameter that is roughly $2 m_{\rm vis}^2 \div 4 m_{\rm vis}^2 $. Since both can be varied over a significant part of this parameter space without violating existing constraints, no fine tuning is required. However the values of $\Delta N_{\rm eff}$ obtained can be close to existing limits so a signal in future experiments is possible.

A strong constraint on the present model comes from requiring that the energy in the visible, i.e. the most strongly reheated, sector dominates the universe before BBN. This happens provided sufficient time passes between particle production and the visible sector temperature dropping to $\sim \MeV$ for the inflaton kinetic energy to redshift enough.
Taking the energy density in the visible sector immediately after reheating to be $E_{0}$, the energy densities in the inflaton and visible sectors at later times are $\frac{1}{2}\dot{\phi}_0^2  \left(a_0/ a\left(t\right)\right)^6$ and $E_0 \left(a_0/a\left(t\right)\right)^4$ respectively, where $a_0$ is the scale factor of the universe at reheating and $\dot{\phi}_0$ is the inflaton speed at this time (previously just called $\dot{\phi}$). Therefore these two sectors have the same energy density when
\beq \label{eq:a0eq}
\left(\frac{a\left(t\right)}{a_0}\right)^2 = \frac{\dot{\phi}_0^2}{2 E_{0}} ~.
\eeq
 The visible sector temperature is at least $T_{\rm min}$ when Eq.~\eqref{eq:a0eq} is satisfied if
\beq \label{eq:rneed}
\frac{E_{0}}{\Lambda^4} \gtrsim  \left(\frac{\pi^2}{30} g_{\rm vis} \frac{T_{\rm min}^4 \dot{\phi}_0^4}{4 \Lambda^{12}} \right)^{1/3} ~,
\eeq
where $g_{\rm vis}$ is the number of visible sector effective relativistic degrees of freedom, and $g_{\rm vis} \simeq 10$ for $T_{\rm min} \sim \MeV$.

In our simple model, avoiding an exponential suppression of the energy transfer (which would make Eq.~\eqref{eq:rneed} much harder to satisfy) requires $g^2 \dot{\phi}_0 \gtrsim \left(100~\GeV \right)^2$. Additionally the form of the inflaton couplings imposes $ \dot{\phi}_0^2 \gtrsim N g^2 \Lambda^2$ (discussed below Eq.~\eqref{eq:infcou}), and for the Higgs masses of the different sectors to be spaced closely enough to avoid tuning $\Lambda^2/N \sim \left(100~\GeV \right)^2$. Combining these,  $\dot{\phi}_0 \sim \Lambda^2$  leading to a lower bound on $E_0$ from Eq.~\eqref{eq:rneed}. Numerical investigation shows that the energy transfer for the most efficient value of $m_h^2$ scales approximately as $g^{4}$. Therefore in this model there is an upper bound on $\Lambda$ for sufficient reheating. Taking $\Lambda \sim 10~\TeV$, with the previously plotted parameter values $g = 0.02$ and $\dot{\phi}_0 = \Lambda^2/\sqrt{8}$ so that the maximum energy transfer happens at approximately the correct Higgs mass, leads to $E_0/\Lambda^4 \gtrsim 3 \times 10^{-10}$ for radiation domination at a temperature of $2~\MeV$. This is likely to already be in slight tension with measurements, and from Figure~\ref{fig:2} is only just reached in this model, although it is not necessary to actually fine tune onto the top of the peak in energy transfer since $-0.5 m_{\rm peak}^2 \lesssim -m_h^2 \lesssim -1.7 m_{\rm peak}^2$ satisfies this bound.

However, we do not have a complete model of inflation, and have made very simple assumptions about the inflaton's  subsequent evolution and couplings, for example in taking its speed constant. We have also not included the possibility of reheating to fermions or gauge bosons or possible effects from inflaton Higgs mixing. As a result, while the general picture of not being able to raise the cutoff to more than intermediate scales is likely to remain, it is at least plausible that radiation domination at temperatures safely above BBN is possible in a complete model. 

An inflaton potential and couplings for which $\dot{\phi}_0$ could be smaller relative to $\Lambda^2$ would lead to a higher temperature when radiation domination begins, due to the decrease in the required $E_0$ from Eq.~\eqref{eq:rneed}, and different inflaton couplings could also lead to more efficient energy transfer. For example, with a pure exponential coupling the occupation of a Higgs mode after reheating, given by Eq.~\eqref{Eq:nnk}, can be $n_k > 1$ in parts of parameter space, whereas the potential we have studied has $n_k < 1$ similar to a quartic coupling, Eq.~\eqref{eq:nk}. It might even be possible to find a model in which there are several moments of particle production, before the inflaton potential becomes flat. For example this could happen with an axion-like periodic coupling between the inflaton and the Higgs, which could exponentially increase the energy transfer due to a Bose enhancement.

The current model also requires a coincidence of scale, similar to the $\mu$ problem in supersymmetry. For radiation domination before BBN, the visible sector Higgs mass must be in a part of parameter space where the energy transfer has a fairly weak dependence on its Higgs mass, i.e. the region $-g^2 \Lambda^2 < m_h^2 <0$ in Figure~\ref{fig:2}. However to avoid constraints from $\Delta N_{\rm eff}$ requires that the energy transfer to the sector with the next lightest Higgs mass is fairly strongly suppressed. Therefore the visible sector Higgs mass squared, which is approximately $-\Lambda^2/N$, must be the same order of magnitude as the unrelated scale $g \dot{\phi}$ at which the adiabatic suppression becomes relevant. This would be relaxed if particle production was more efficient, for example due to a different form of the inflaton couplings. In such a model, the visible sector could be in a part of parameter space where the energy transferred has a strong dependence on the Higgs mass squared, rather than having to coincide closely with the unrelated adiabaticity scale.\footnote{A peak where the inflaton and bare Higgs mass cancel in a part of parameter space where production is otherwise exponentially suppressed could also help, although this would still require a coincidence that the width of the peak must be comparable to the spacing of the Higgs masses.}

\section{Discussion}\label{sec:5}

In this paper we have studied non-perturbative reheating in models in which the inflaton does not oscillate around a minimum of its potential but instead runs away towards infinity after inflation. In these there are no perturbative inflaton decays and non-perturbative processes determine the final energy transferred to a sector. In particular, we have considered particle production with more general inflaton couplings than previous work \cite{Felder:1999pv}, and also the possibility of this happening after low scale inflation. As a result we have shown that such models have dynamics that can realise the Nnaturalness solution to the hierarchy problem. The energy transferred to a sector can be exponentially dependent on the mass of its Higgs, and potentially dangerous constraints from the contributions of the other sectors to cosmological observables are easily accommodated. Another significant difference to the original Nnaturalness model is that the mass squared of the Higgs in a sector is compared to the scale $\dot{m}_h$, which depends on a coupling constant, rather than the mass of a reheaton (although it can be technically natural for the later to be small).

However, the present implementation requires some features that might be challenging to obtain from a UV theory. The inflaton needs to decouple from the $N$ sectors at late times, and it must also have a flat potential at large field values. A major deficiency of our work is the absence of an explicit example of dynamics leading to such properties. Further we are far from having a complete model of inflation, and have instead focused on the features of reheating. Additionally, we have neglected possible effects from production of gauge bosons and fermions, and in our simplified model reheating sufficiently that radiation domination begins before BBN is challenging and the cutoff cannot be raised above $\sim 10~\TeV$. We hope to return to these issues in future work.

Another downside is that observational signatures from mixing between sectors are not automatically present since the inflaton decouples (unlike the original Nnaturalness model \cite{Arkani-Hamed:2016rle}).
 Despite this the sector with the next lightest Higgs can still give a significant contribution to $\Delta N_{\rm eff}$ which could be detected in the future. Additionally, the cosmological evolution is non-standard with the universe dominated by the inflaton kinetic energy until low temperatures (in the present model, almost until BBN). This could be observable through deviations compared to the standard BBN predictions, or via its effect on the dark matter relic abundance \cite{Salati:2002md,Profumo:2003hq,Rehagen:2014vna,Redmond:2017tja}. The scale of inflation is also necessarily low, and a mechanism for producing a baryon asymmetry consistent with this must be realised (e.g. \cite{Hambye:2004jf,Pilaftsis:2005rv}). The discovery of new physics allowing this would at least be compatible with the present model, and optimistically a measurement of such a theory's numerical parameters might even give a hint towards a period of kinetic energy domination. Finally, at least in the present model, above $10~\TeV$ some other solution to the hierarchy problem is required, and this could be be observed in a future collider \cite{Arkani-Hamed:2015vfh}.
 
In passing we also note that these dynamics allow a potential, albeit very speculative, half-way path between  anthropic and dynamical explanations of the EW hierarchy problem (with some of the advantages and disadvantages of both). It might be that a light Higgs is not directly required for the development of life. Instead it could be needed so that the universe is reheated to a sufficiently high temperature that some other process could occur, which itself is anthropically necessary, for example leptogenesis. If disconnected regions of the universe have different Higgs masses this could provide an explanation for the hierarchy problem. Although extremely hard to test in practice, evidence of a reheating process that is dependent on the Higgs mass is in principle observable, and such a scenario would also automatically evade constraints from $\Delta N_{\rm eff}$. Additionally, a non-oscillating inflationary model is not necessary since perturbative inflaton decays lead to a much lower maximum temperature than preheating.

Similarly speculatively, models of Nnaturalness do not necessarily face the challenge from technicolour that is potentially worrying for pure anthropic models (mentioned in the Introduction). Even if the $N$ sectors include a significant number that realise EW symmetry breaking with a new strongly coupled gauge group, the UV theory may well be such that these do not get reheated significantly. For example reheating to technicolour sectors containing only chiral fermions is likely to be strongly suppressed. UV theories containing light vector-like fermions could plausibly also have suppressed couplings to the inflaton, for example if it has an approximate $Z_2$ symmetry.

\subsection*{Acknowledgements}
I am grateful to the Mainz Institute for Theoretical Physics
(MITP) for its hospitality during the completion of this
work, and to Nayara Fonseca for very useful comments on a draft.

\bibliographystyle{JHEP}
\bibliography{reference}

\end{document}